\documentclass{sf2a-conf2017}
\usepackage{graphicx}
\usepackage{hyperref}
\usepackage[]{natbib}  
\usepackage{epstopdf}

\def\BibTeX{{\rm B\kern-.05em{\sc i\kern-.025em b}\kern-.08em
    T\kern-.1667em\lower.7ex\hbox{E}\kern-.125emX}}
\bibpunct{(}{)}{;}{a}{}{,}  %%%%%%%%%%%%%  A&A bibliography style
\begin{document}

\TitreGlobal{SF2A 2017}

%%-----------------------------------------------------------------
%%      the top matter
%%

\title{Does the existence of a plane of satellites \\ constrain properties of the Milky Way?}

\runningtitle{Correlated satellite structures and MW properties}

\author{Marcel S. Pawlowski}\address{Hubble Fellow, Department of Physics and Astronomy, University of California, Irvine, CA 92697, USA}

%% Keep this line, even if the page will be settled afterwards.
\setcounter{page}{237}

%%-----------------------------------------------------------------

\maketitle

%%-----------------------------------------------------------------
%%        The abstract
%% 
%%  Warning!  within the abstract:
%%  - do not use macros. 
%%  - do not use commands like: \cite, \citet, \citep ... etc.

\begin{abstract}
According to the hierarchical model of galaxy formation underlying our current understanding of cosmology, the Milky Way (MW) has continued to accrete smaller-sized dwarf galaxies since its formation. Remnants of this process surround the MW as debris streams and satellite galaxies, and provide information that is complementary to studies of the Galaxy itself. The satellite system thus has the potential to teach us about the formation and evolution of the MW. Can the existence of a narrow, co-rotating plane of satellite galaxies (the Vast Polar Structure, VPOS) put constraints on our Galaxy's properties? Are such satellite galaxy planes more narrow around less massive hosts, more abundant around more concentrated hosts, more kinematically coherent around more early-forming halos? To address such questions, we have looked for correlations between properties of satellite galaxy planes fitted to cosmological simulations in the ELVIS suite and properties of their host dark matter halos, while accounting for realistic observational biases such as the obscuration by the disk of the MW. We find no evidence for strong correlations that would allow conclusions on the host halo properties from the mere existence of the VPOS around our Galaxy.
\end{abstract}

%% Insert the keywords (to appear in the ADS indexing)
%% Keywords must be separated by a comma

%%-----------------------------------------------------------------

\section{Introduction}
%%---------------------

Both major galaxies in the Local Group host planes of satellite galaxies. The Vast Polar Structure (VPOS) of the Milky Way (MW) has a root-mean-square (rms) height of $\approx 20$\,kpc and extends perpendicular to the disk of the MW until at least its virial radius. Globular clusters and streams of disrupted systems show a preference for a similar orientation \citep{Pawlowski2012}. Proper motion measurements indicate that a majority of the 11 classical satellite galaxies co-orbit along the VPOS \citep{Pawlowski2013}.
Among the satellites of the Andromeda galaxy M31, about 50\% have been identified to be part of a narrow Great Plane of Andromeda (GPoA, \citealt{Ibata2013}). This structure has a rms height of only $\approx 14$\,kpc. Its fortuitous edge-on orientation with respect to the Sun allows to identify a coherent line-of-sight velocity trend: 13 of 15 satellites are consistent with a rotating satellite plane (though tangential motions are unknown and could be high enough to quickly disperse the structure, \citealt{Gillet2015,Buck2016}).
Furthermore, there is increasing evidence for similar correlations in more distant satellite systems (\citealt{Chiboucas2013,Ibata2014,Tully2015,Mueller2016}, Mueller et al. in prep).

Comparisons of satellite galaxy planes with sub-halo systems in cosmological simulations based on the $\Lambda$CDM model show that similarly anisotropic and kinematically correlated arrangements are very rare \citep{Ibata2014b,Pawlowski2014b,Pawlowski2014}. While most of these simulations are dark-matter-only, the existence of pronounced planes of satellite galaxies poses a fundamental problem for which even the inclusion of baryonic physics does not offer an obvious solution \citep{Pawlowski2015,Ahmed2017,Pawlowski2017}. 

However, it is in principle imaginable that only a subset of halos with specific properties can host pronounced satellite planes. If this were the case, the observation of the VPOS could constrain properties the MW. \citet{Buck2015} have claimed to have found evidence that the existence of narrow, kinematically coherent satellite planes comparable to that of M31 is linked to properties of the host halo. In their analysis of 21 simulated satellite systems they see a correlation between the concentration parameter of the host halo and satellite planes, in the sense that more concentrated halos contain more narrow planes of satellites. They argue that halo concentration acts as a proxy for formation time: later-forming halos are less concentrated and thus do not contain as pronounced satellite planes.

Their analysis was confined to analogs of the M31 system, and did not take observational biases and uncertainties into account (such as the PAndAS survey footprint within which most known M31 satellites were discovered). Instead, they only selected satellites from within the virial volume. Since their argument was based on the M31 satellite plane, which consists of only a sub-sample of all M31 satellites, it is unclear whether such a signal persists for analogs of the MW system. To investigate this, we have analyzed the ELVIS simulations and compared them to the VPOS, more specifically to the positional and kinematic correlations found among the 11 classical MW satellite galaxies. This analysis is based on \citet{Pawlowski2014}, but now investigates whether the planes fitted to the sub-halo satellite systems in those simulations show any correlations with properties if the host halo. A more detailed comparison to the GPoA will be presented in a future publication.

\section{Satellite plane fits in ELVIS: are there correlations with host halo properties?}
%%-------------------------

\citet{Pawlowski2014} focussed on testing for an environmental dependency of the occurrence and properties of planes of satellite galaxies. They found no pronounced differences between hosts that are isolated and hosts that are in a paired configuration similar to the MW-M31 system. The study used the Exploring the Local Volume in Simulations (ELVIS) suite of cosmological, dark-matter only simulations by \citet{Garrison-Kimmel2014}. All 48 host halos in ELVIS are used, and the 11 top-ranked satellites (by peak mass) are selected as the equivalent of the 11 brightest, ``classical'' satellite galaxies of the MW (which are also the only ones for which proper motion measurements provide information on their orbital coherence). The effects of uneven sky coverage due to obscuration by the disk of the MW is modeled by drawing 100 realizations with randomly oriented MW disks that obscure 20\,\% of the sky. More details on the satellite selection and plane fitting can be found in \citet{Pawlowski2014}.

Using this dataset, we investigate the possibility of a dependence between properties of the host halos and properties of planes fitted to the satellite distribution. Specifically, we consider three measures of satellite plane coherence, averaged over the 100 realizations per host: 
(1) $r_\mathrm{per}$, the rms height of satellites relative to the best-fit plane (an absolute measure of the flattening),
(2) $c/a$, the rms short-to-long axis ratio of the satellite distribution (a relative measure of the flattening), and
(3) $\Delta_\mathrm{std}$, the spherical standard deviation of the eight closest-clustering satellite orbital poles (a measure of the orbital coherence).

\begin{figure}[ht!]
 \centering
 \includegraphics[width=0.33\textwidth,clip]{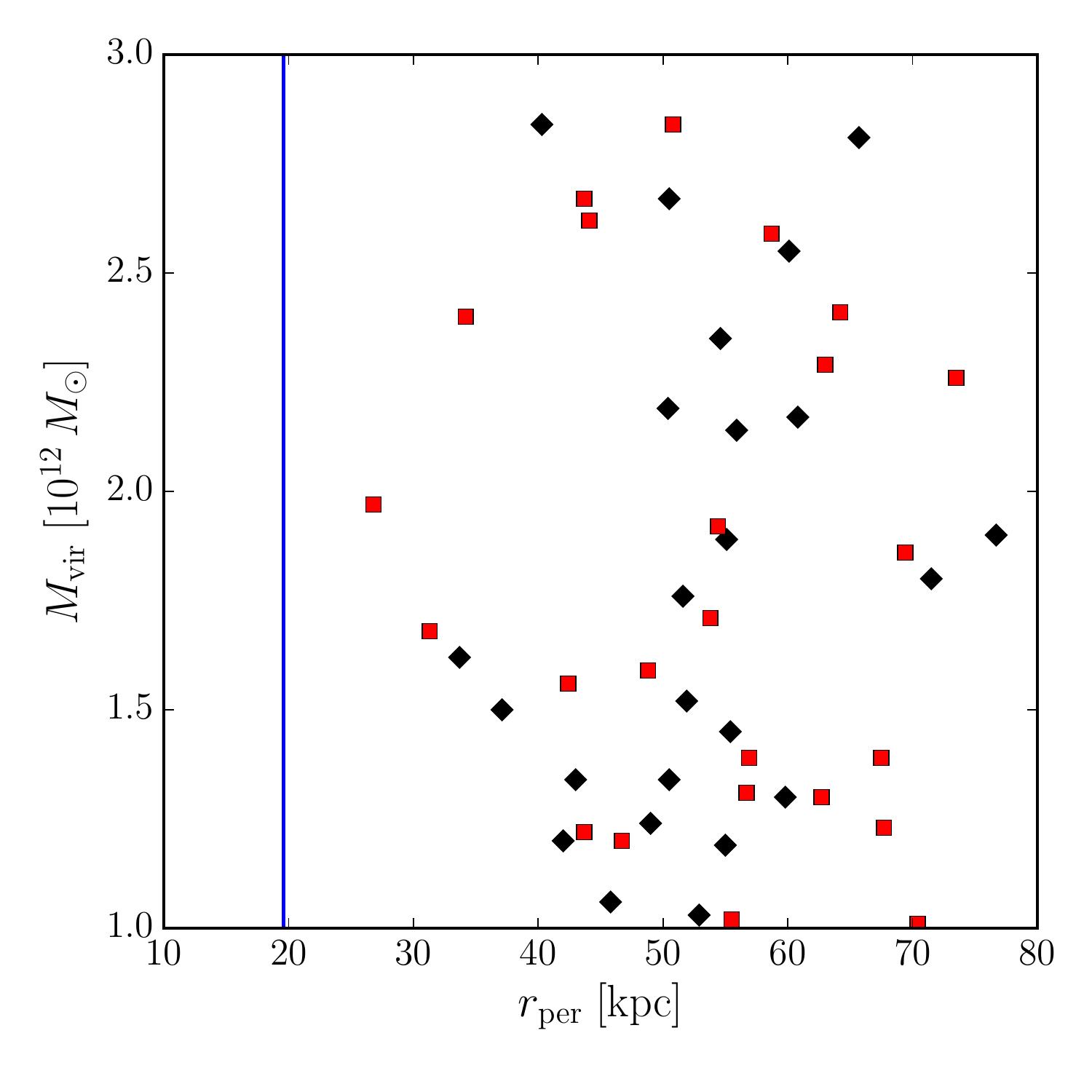}% 
 \includegraphics[width=0.33\textwidth,clip]{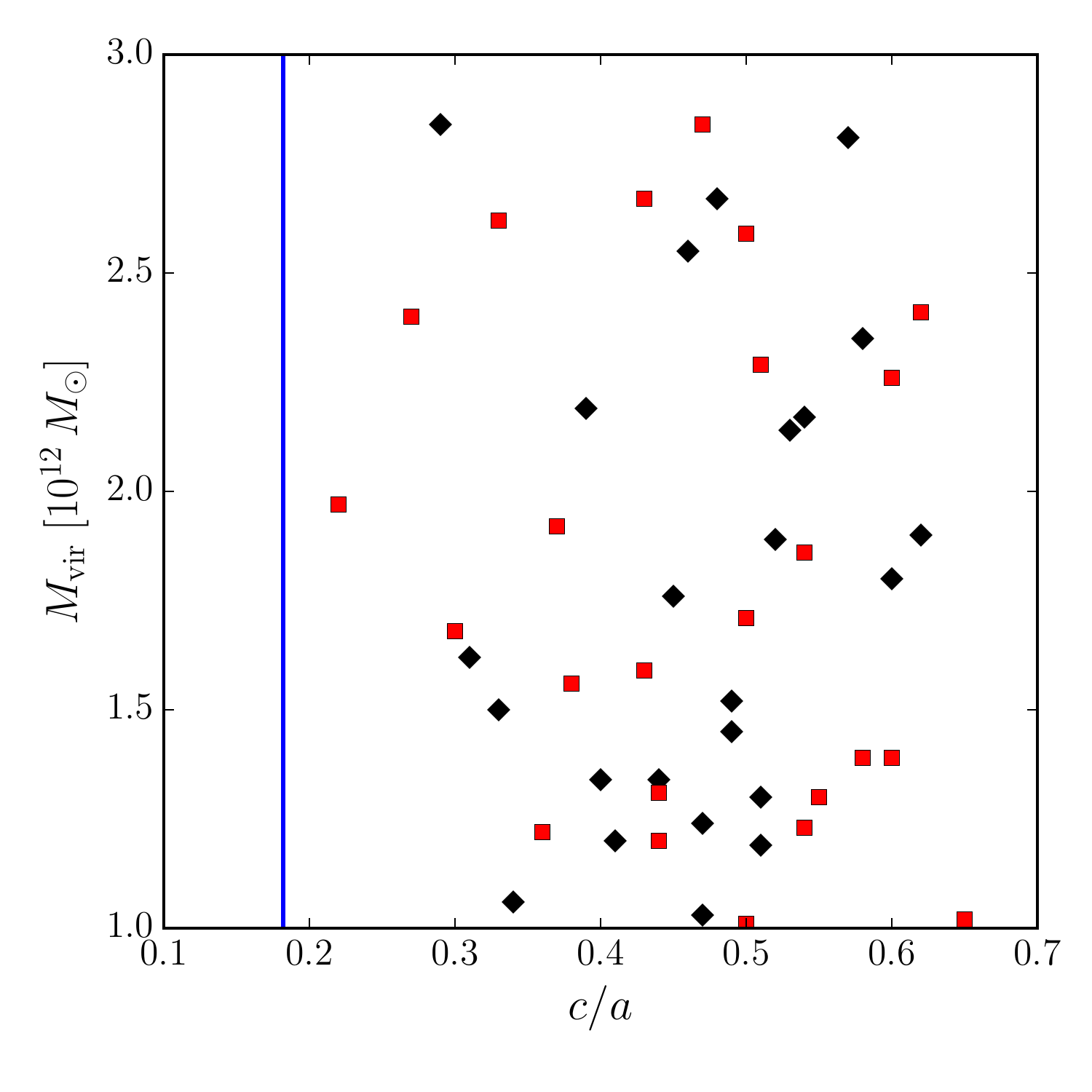}%      
 \includegraphics[width=0.33\textwidth,clip]{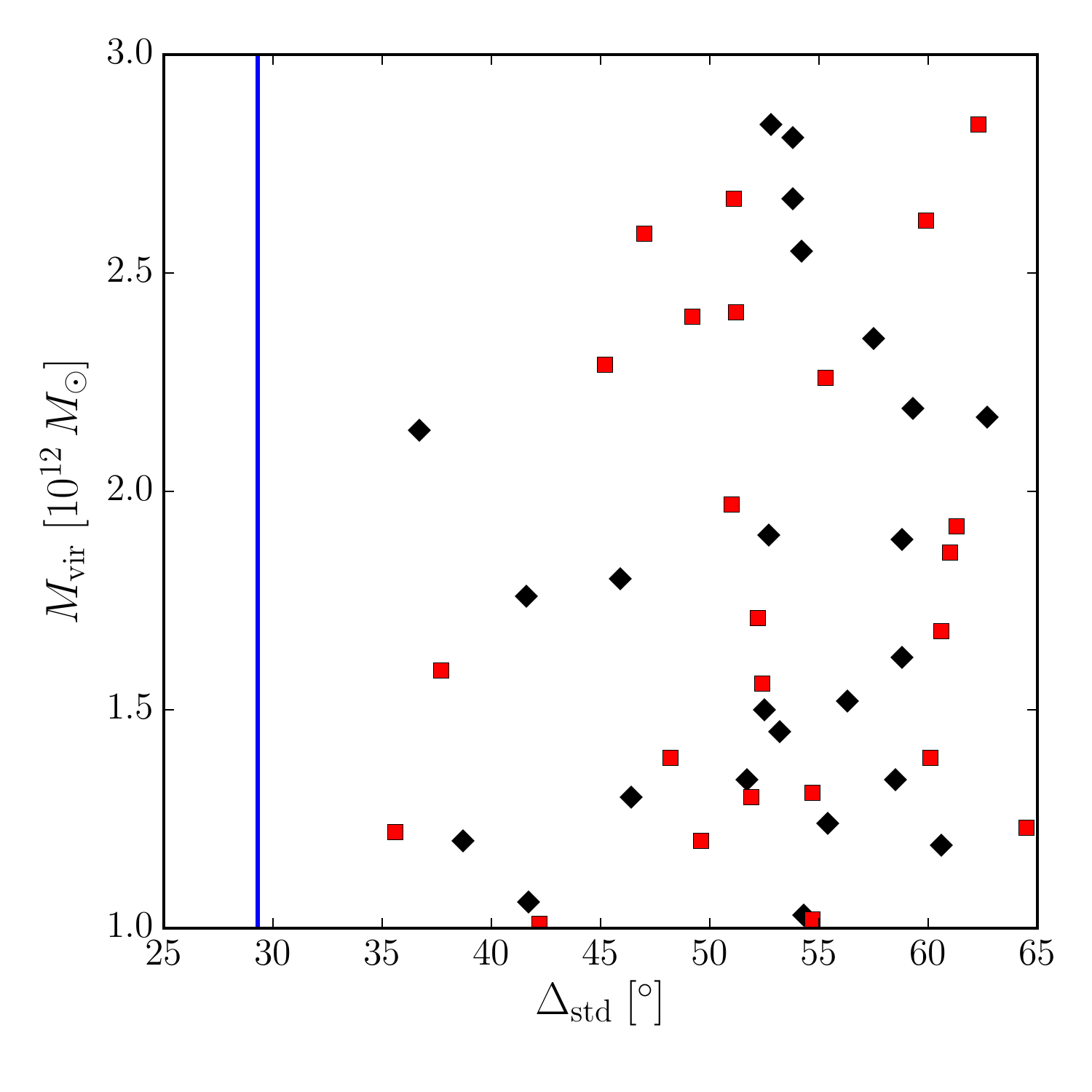}      
%% Note the ABSENCE of the extension .pdf  !
  \caption{Host halo virial mass $M_{\mathrm{vir}}$\ plotted against rms height $r_\mathrm{per}$ ({\bf left}), the short-to-long axis ratio $c/a$\ ({\bf middle}), and the spherical standard deviation of the eight best-aligned sub-halo orbital poles ({\bf right}) of the 11 top-ranked sub-halos. Shown are the average values for the sub-halos systems corresponding to the 48 host halos in the ELVIS suite of cosmological simulations. Isolated host halos are plotted as red squares, host halos in a paired configuration similar to that of the Local Group are plotted as black diamonds. The corresponding value for the observed system of the 11 classical MW satellite galaxies are indicated as blue lines. 
  }
  \label{pawlowski1:fig1}
\end{figure}

\begin{figure}[ht!]
 \centering
 \includegraphics[width=0.33\textwidth,clip]{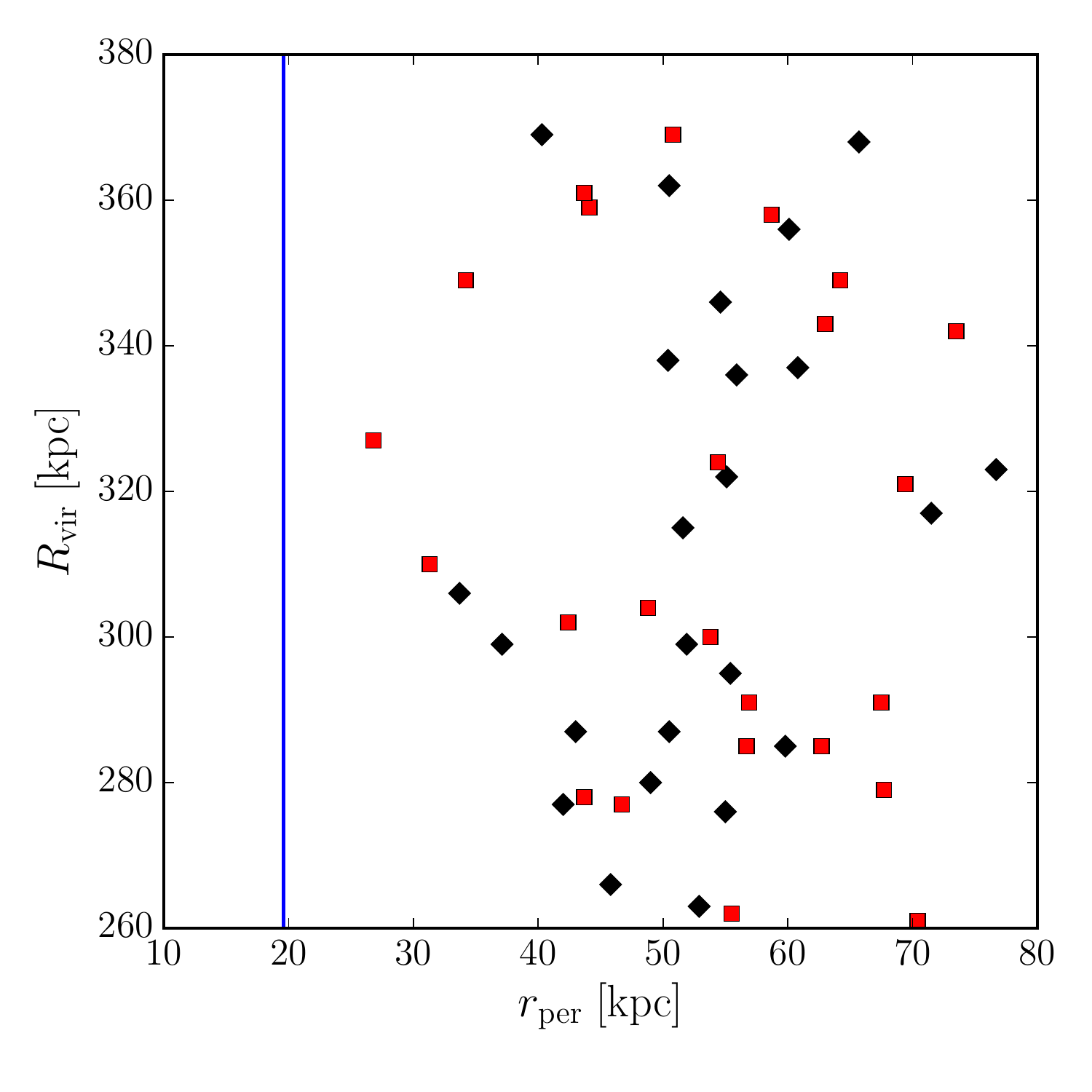}%      
 \includegraphics[width=0.33\textwidth,clip]{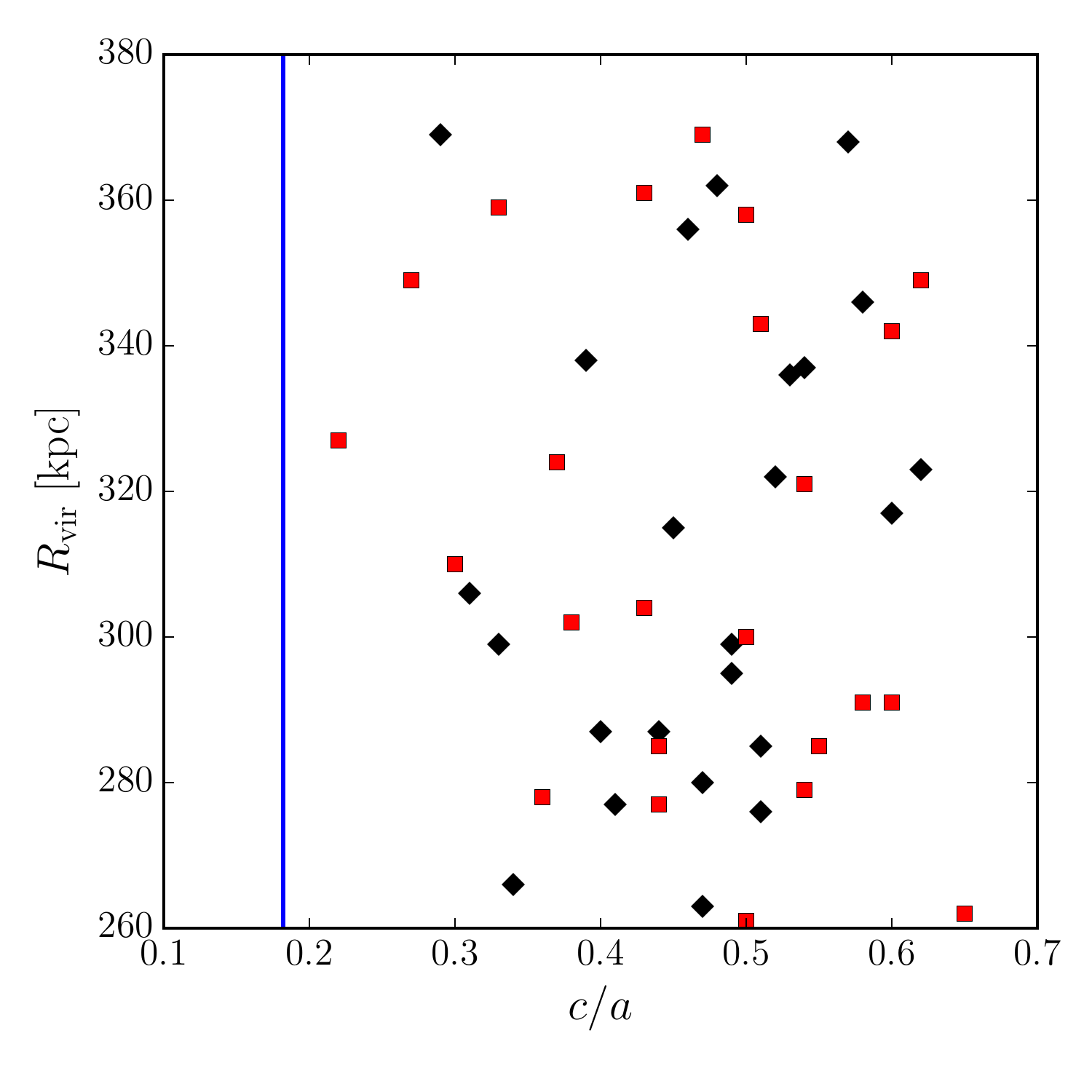}      
 \includegraphics[width=0.33\textwidth,clip]{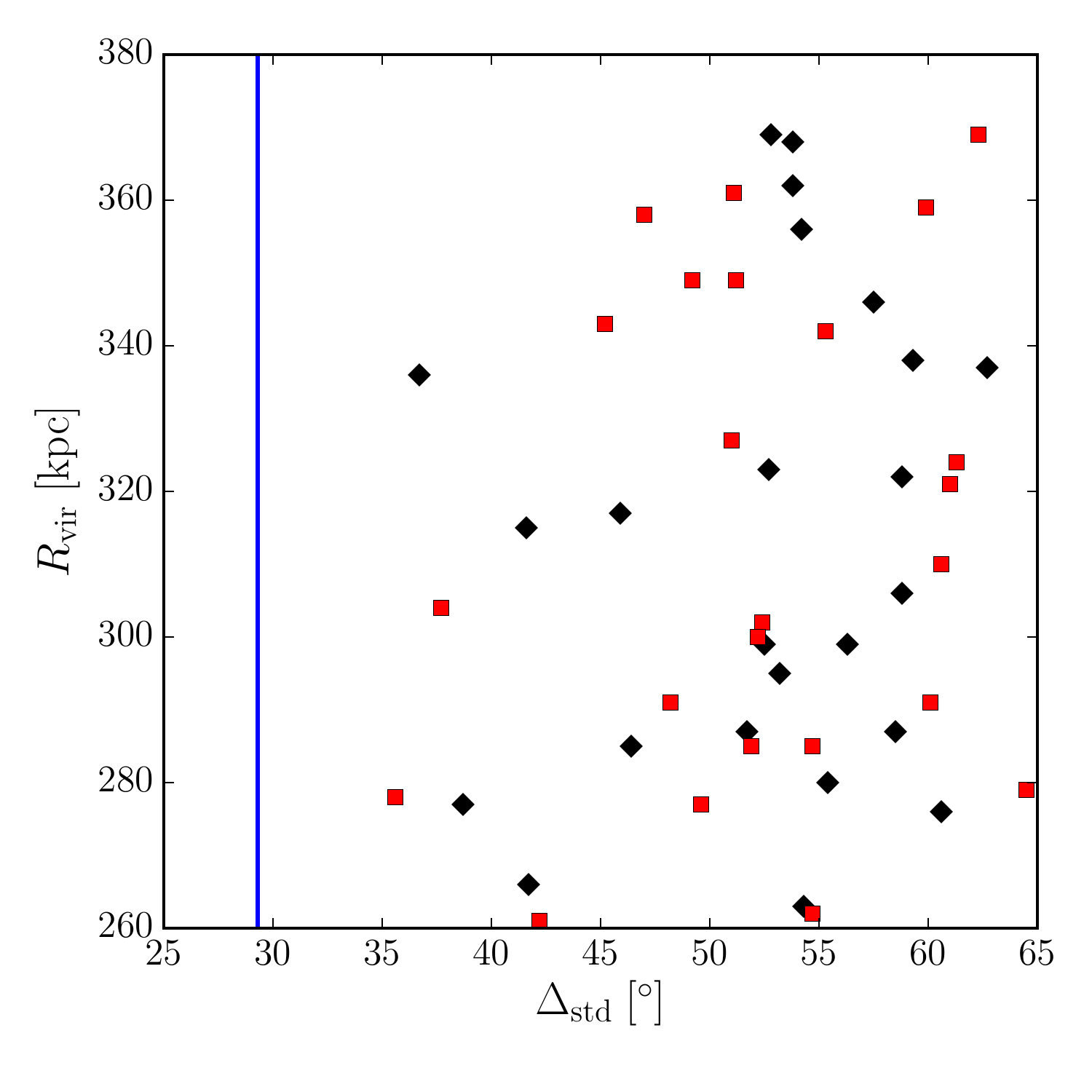}      
%% Note the ABSENCE of the extension .pdf  !
  \caption{Same as Figure \ref{pawlowski1:fig1}, but for the host halo virial radius $R_{\mathrm{vir}}$.}
  \label{pawlowski1:fig2}
\end{figure}

\begin{figure}[ht!]
 \centering
 \includegraphics[width=0.33\textwidth,clip]{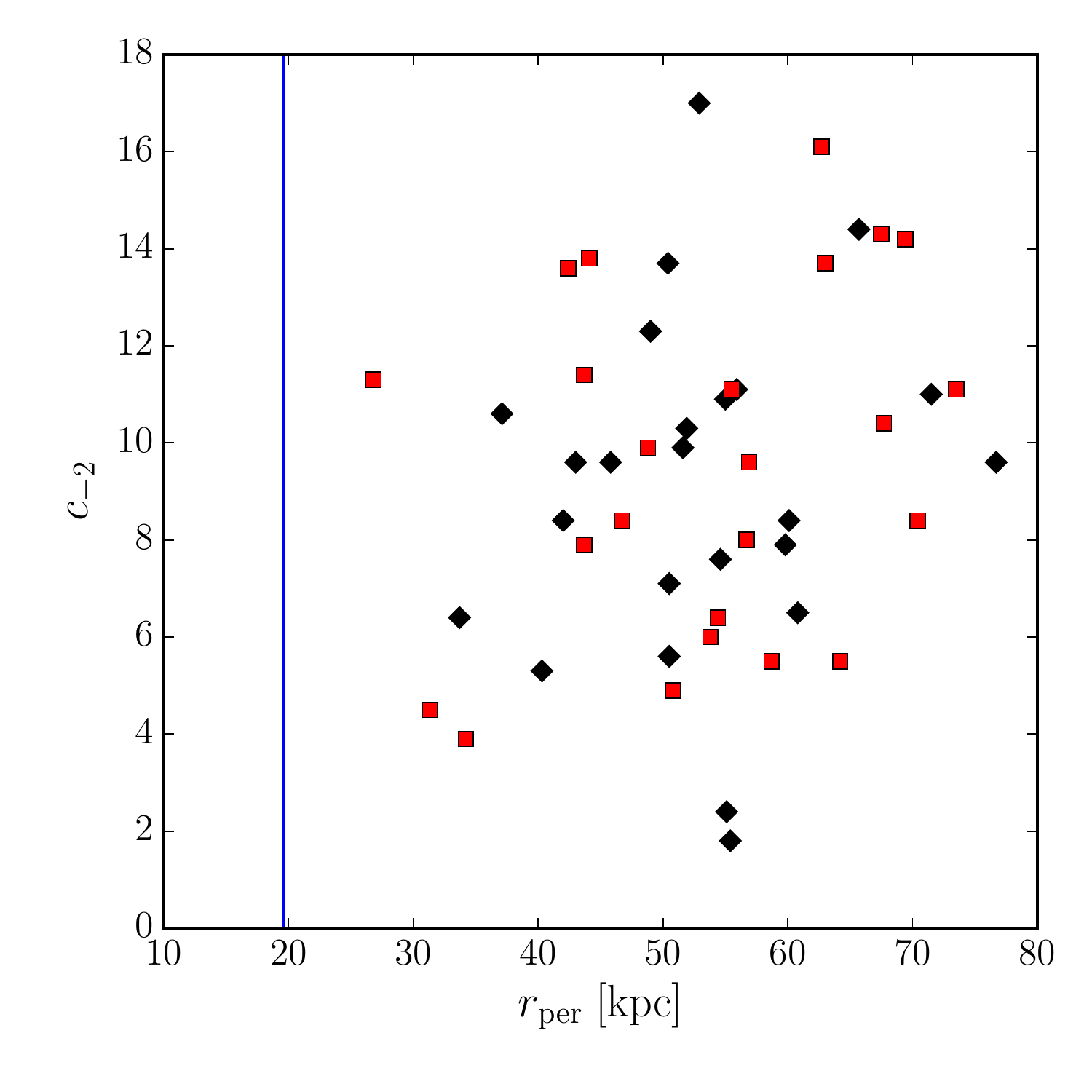}%      
 \includegraphics[width=0.33\textwidth,clip]{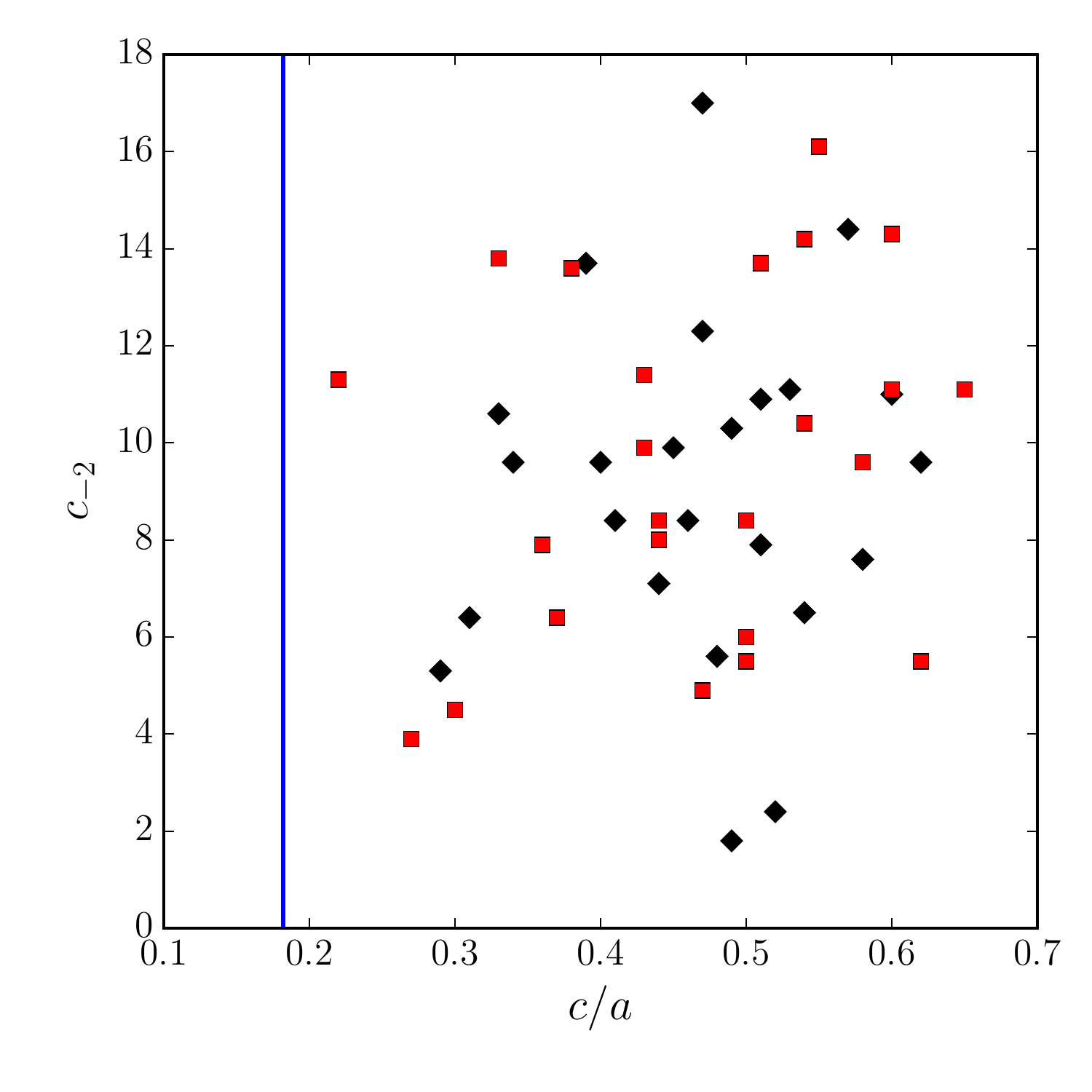}%      
 \includegraphics[width=0.33\textwidth,clip]{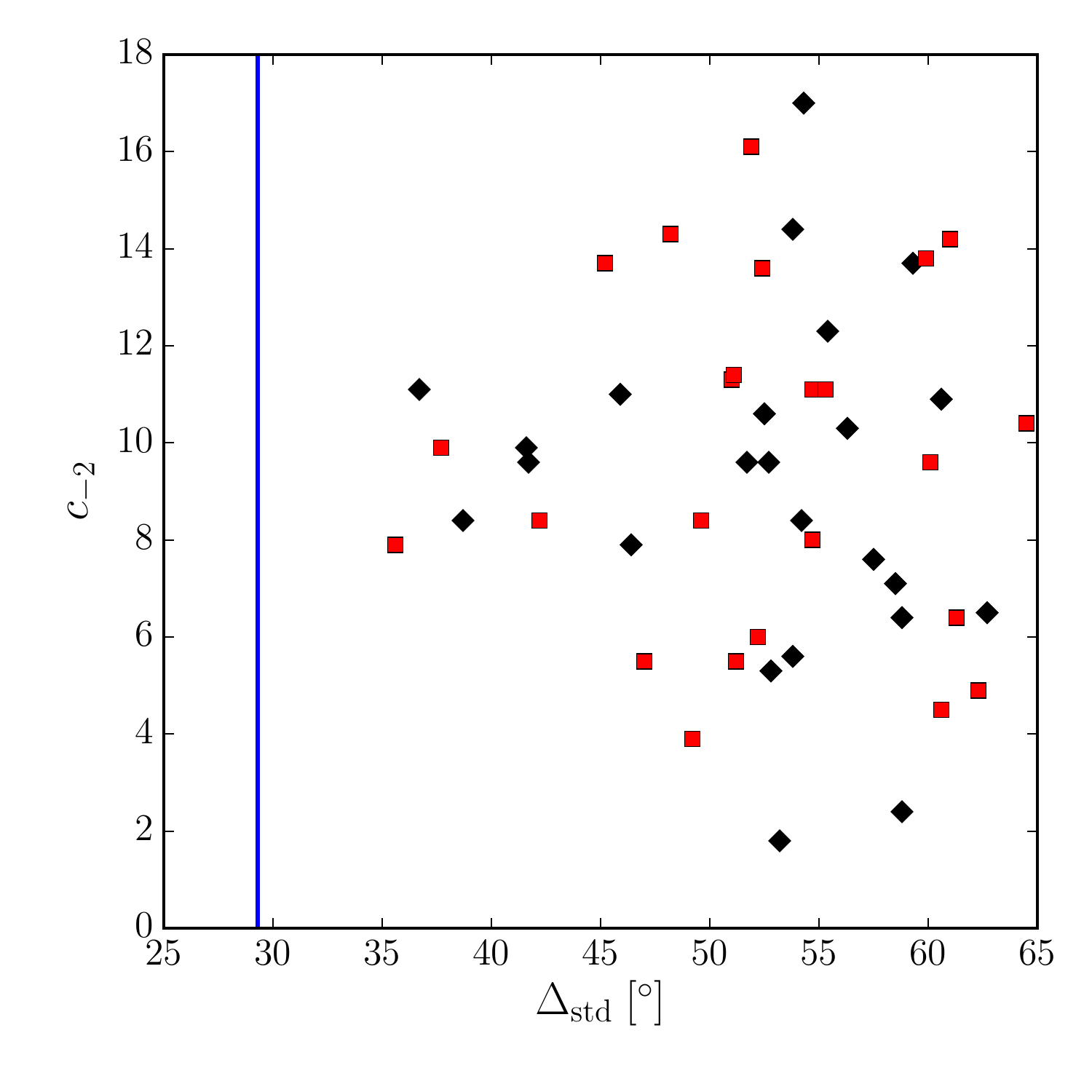}      
%% Note the ABSENCE of the extension .pdf  !
  \caption{Same as Figure \ref{pawlowski1:fig1}, but for the host halo concentration parameter $c_{\mathrm{-2}}$.}
  \label{pawlowski1:fig3}
\end{figure}

\begin{figure}[ht!]
 \centering
 \includegraphics[width=0.33\textwidth,clip]{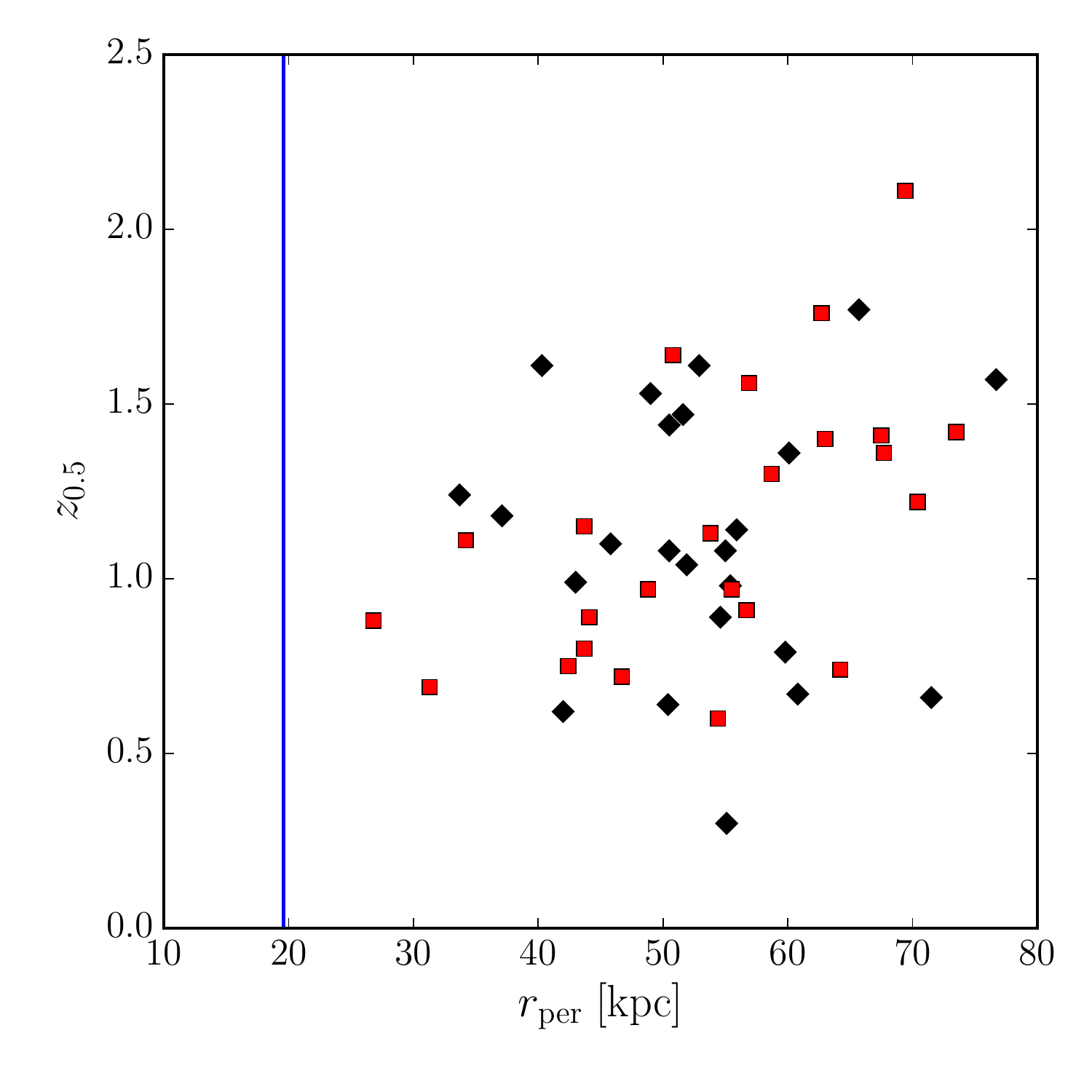}% 
 \includegraphics[width=0.33\textwidth,clip]{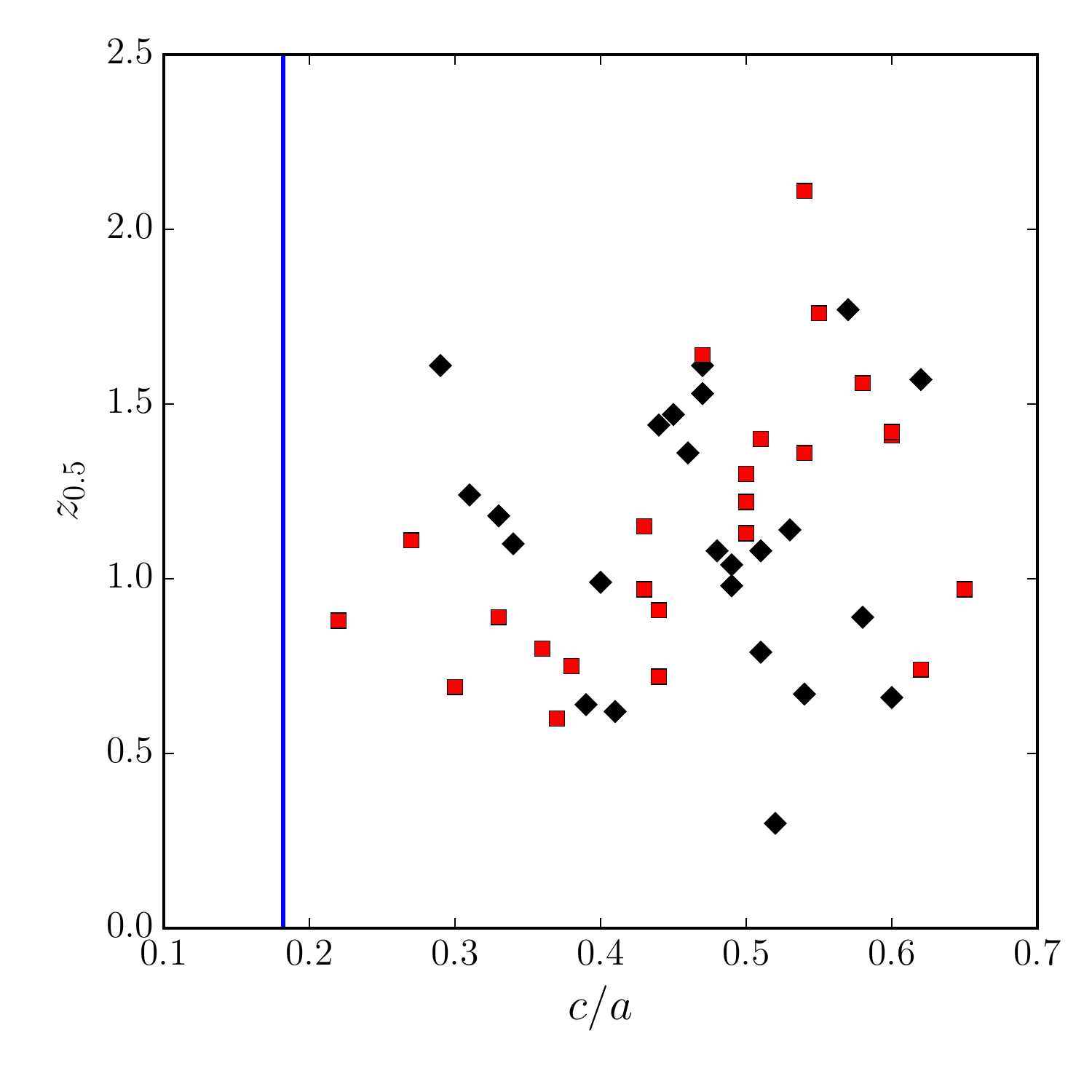}%      
 \includegraphics[width=0.33\textwidth,clip]{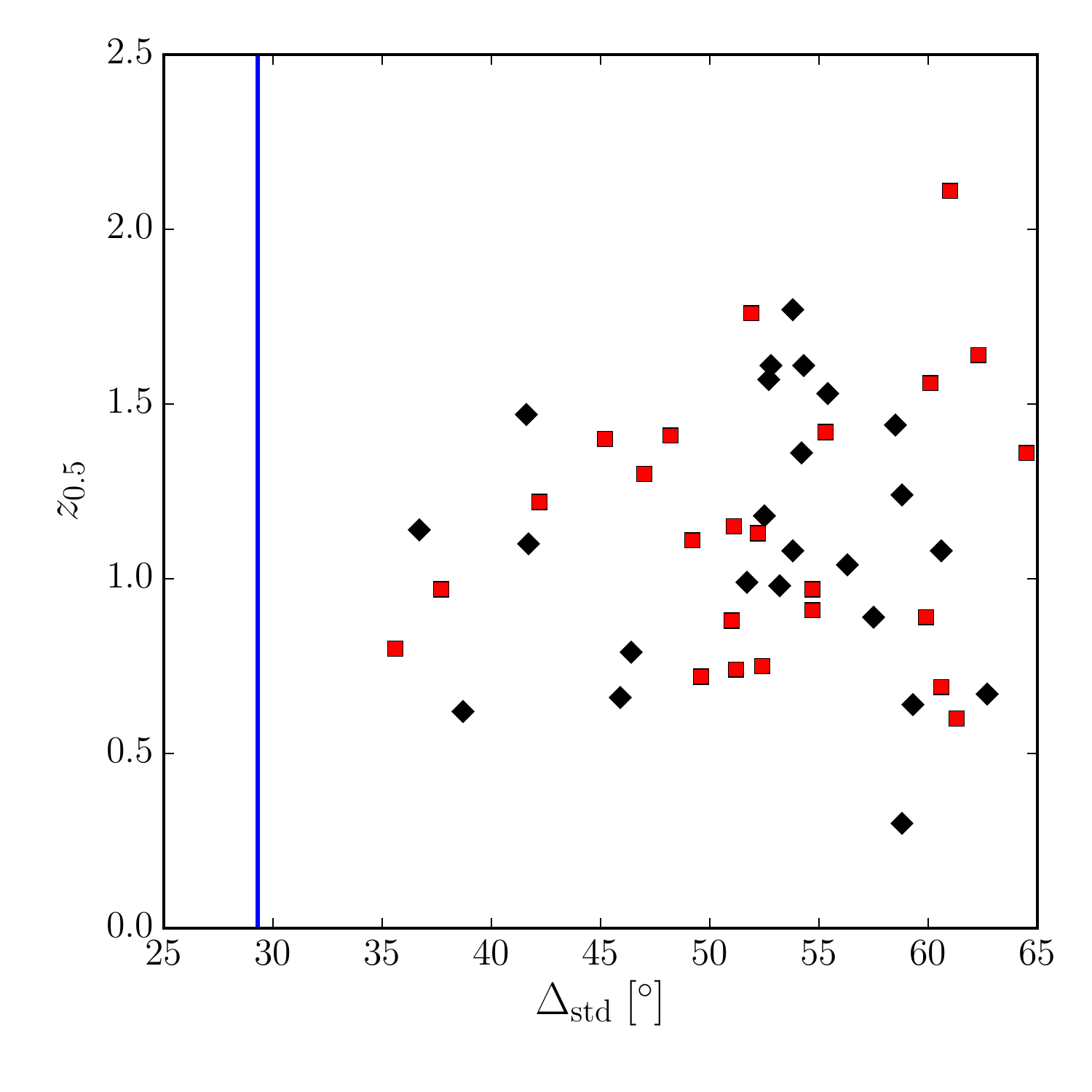}      
%% Note the ABSENCE of the extension .pdf  !
  \caption{Same as Figure \ref{pawlowski1:fig1}, but for the host halo formation redshift $z_{\mathrm{0.5}}$.}
  \label{pawlowski1:fig4}
\end{figure}

Figures \ref{pawlowski1:fig1} to \ref{pawlowski1:fig4} plot these three measures against four properties of the host halos: 
(1) $M_\mathrm{vir}$, the virial mass,
(2) $R_\mathrm{vir}$, the virial radius,
(3) $c_{-2}$, the halo concentration (equivalent to an NFW halo concentration), and 
(4) $z_\mathrm{0.5}$, the formation redshift, defined as that redshift $z$\ where the progenitor halo first reached half of the final host halo mass.

As in \citet{Pawlowski2014}, each symbol in the figures represents one of the ELVIS host halos and is coded for whether it is isolated (red square) or part of a pair (black diamonds). Inspecting the figures shows that none of the 48 host halos typically contain satellite planes with properties as extreme as those observed for the 11 classical satellites (blue lines). Simulated satellite systems are wider, less flattened, and less kinematically coherent than the observed VPOS. The plots also show no difference in the plane coherence parameters between the isolated and the paired host halos, in line with the conclusion of \citet{Pawlowski2014}.

Most importantly, the plots demonstrate that there are no clear correlations between properties of the host halo and properties of planes fitted to their satellite system. The only possible weak trend is with the halo formation redshift $z_{0.5}$: the most narrow ($30 \leq r_\mathrm{per} \leq 45\,\mathrm{kpc}$), most flattened ($0.2 \leq c/a \leq 0.4$), and most kinematically correlated ($35^\circ \leq \Delta_\mathrm{std} \leq 50^\circ$) planes tend to be preferentially found for later forming halos with $0.6 \leq z_{0.5} \leq 1.3$. Other host halos with similar $z_{0.5}$\ do not contain as correlated satellite planes, but the earlier forming hosts in ELVIS appear to avoid these regions of plane coherence parameters.

\section{Discussion and Conclusion}
%%--------------------

Intriguingly, the tendency to find slightly more correlated satellite systems around later forming hosts is the {\it opposite} of the trend reported by \citet{Buck2015}. It is also in tension with the formation history of the MW, which is believed to not have experienced a major merger since $z \approx 2$. As such, and in combination with the difficulty of even finding any simulated satellite systems that resemble the strong coherence found for the VPOS, it appears unlikely that the weak tendency of more correlated satellite systems living in later-forming hosts can provide a reliable constraint on the halo properties of the MW, or even solve the satellite plane problem.

However, structures of satellite galaxies can offer other constraints on their host galaxy properties. If satellite planes are found to be stable, this would require close-to spherical halo and thus constrains halo triaxiallity \citep{Fernando2017}. One proposed explanation for satellite planes is the accretion of many satellites in a common group. Identifying satellite galaxies that were accreted as one group can then constrain the MW potential since they must share similar specific angular momenta and energies. Furthermore, more exotic proposed explanations for the occurrence of apparently co-orbiting planes of satellite galaxies have implications for the history of the host galaxies. For example, if satellite galaxy planes are formed out of Tidal Dwarf Galaxies \citep{Bournaud2006,Kroupa2012}, this would imply a past galaxy encounter involving the host (or happening in the vicinity; \citealt{Hammer2013}).

% Optional acknowledgements
% -------------------------
\begin{acknowledgements}
Support for this work was provided by NASA through Hubble Fellowship grant \#HST-HF2-51379.001-A awarded by the Space Telescope Science Institute, which is operated by the Association of Universities  for  Research  in  Astronomy,  Inc.,  for  NASA,  under  contract  NAS5-26555
\end{acknowledgements}

%%-----------------------------
%%   Bibliography
%%-----------------------------
%%

%% The following lines are required when using BibTEX (strongly encouraged!):
\bibliographystyle{aa}  % A&A bibliography style file (aa.bst)
\bibliography{pawlowski1} % your references in file: Yourfile.bib

\end{document}